\documentclass[a4paper]{jpconf}
\usepackage{graphicx}

\usepackage[sort&compress,numbers]{natbib}
\usepackage{hyperref}
\usepackage{color}

\newcommand{\amrex}{{\sf AMReX}}
\newcommand{\microphysics}{{\sf Microphysics}}
\newcommand{\pynucastro}{{\sf pynucastro}}
\newcommand{\castro}{{\sf Castro}}
\newcommand{\maestroex}{{\sf MAESTROeX}}
\newcommand{\nyx}{{\sf Nyx}}
\newcommand{\flash}{{\sf Flash}}

\usepackage{bm}

\newcommand{\Uc}{{\,\bm{\mathcal{U}}}}

\newcommand{\Rb}{{\bf R}}

\newcommand{\Adv}[1]{{\left [\boldsymbol{\mathcal{A}} \left(#1\right)\right]}}

\newcommand{\Advs}[1]{\boldsymbol{\mathcal{A}} \left(#1\right)}

\begin{document}
\title{The Challenges of Modeling Astrophysical Reacting
Flows}

\author{Michael Zingale$^{1}$, Khanak Bhargava$^{1}$, Ryan Brady$^{1}$, Zhi Chen$^{1}$,
        Simon Guichandut$^{2}$, Eric T. Johnson$^{1}$, Max Katz$^{1}$, Alexander Smith Clark$^{1}$}

\address{$^1$ Department of Physics and Astronomy, Stony Brook University, Stony Brook, NY 11794-3800, USA}
\address{$^2$ Department of Physics and Trottier Space Institute, McGill University, 3600 rue University, Montreal, QC, H3A 2T8, Canada}

\ead{michael.zingale@stonybrook.edu}

\begin{abstract}
Stellar evolution is driven by the changing composition of a star from
nuclear reactions.  At the late stages of evolution and during
explosive events, the timescale can be short and drive strong
hydrodynamic flows, making simulations of astrophysical reacting
flows challenging.  Over the past decades, the standard approach to
modeling reactions in simulation codes has been operator splitting, using implicit integrators for reactions.  Here we explore some of the assumptions in this standard
approach and describe some techniques for improving the efficiency and
accuracy of astrophysical reacting flows.
 \end{abstract}

\section{Introduction}

Many stellar phenomena are driven by the interplay of nuclear
reactions and hydrodynamics.  During the different phases of stellar
evolution, the behavior of reaction-driven convection, mixing at
composition interfaces, and the feedback between explosive energy
release and hydrodynamic flow all strongly depend on modeling the
reactions accurately.  Thermonuclear explosions in nova depend on
resolving the interface between the white dwarf and accreted layer,
allowing CNO elements to mix and catalyze the burning.  X-ray burst (XRB) flames in
mixed H/He layers require a large number of nuclei to accurately capture
the flame speed.  Type Ia supernova, in all the various progenitor
models, involves burning fronts that intimately connect hydrodynamics
and reactions.  The convective burning shells in the late stages
of massive star evolution involve mixing between the shells
and require large reaction networks to capture the core evolution.
All of these systems
provide challenges for multidimensional models.

Numerical simulations of these environments require algorithms that
strongly couple the reactions and the hydrodynamics.  Operator
splitting methods, commonly used in astrophysics, can break down \cite{sportisse:2000}, and the common workaround of
simply cutting the timestep or disabling the burning (see, e.g.,
\cite{zingale:2001,Fields_2020,rivas:2022}) is not efficient or
accurate.  Instead, we focus on methods that strongly couple hydrodynamics
and reactions, allowing the reactions to “see” the hydrodynamic
flow and respond immediately as the burning proceeds.  Over the past
few years, we have been developing such methods for astrophysical
flows, based on the ideas of spectral deferred corrections, and have
shown that strongly-coupled time-integration methods can be more
efficient than operator splitting on a variety of problems.
Here we
review our algorithmic development for modeling astrophysical reacting
flows, discuss the application to various astrophysical
environments, and describe our success at utilizing GPUs to efficiently model these events.

\section{The \amrex-Astrophysics Suite}

The \amrex-Astrophysics suite \cite{astronum:2017} is a collection of open-source
astrophysics simulation codes built around the \amrex\ adaptive mesh
library \cite{amrex_joss}, and includes the compressible hydrodynamics
code \castro~\cite{castro,castro_joss}, the low Mach number
hydrodynamics code \maestroex~\cite{maestroex}, and the cosmology code
\nyx~\cite{nyx}.  These codes are designed for exascale computers and to be
highly performance-portable, making excellent use of GPUs
\cite{castro_gpu}.  Our approach to using GPUs is to allocate the
state data directly on the GPU at the start of the simulation and then
run all of the physics on the GPUs.  To enable this, we utilize the
AMReX C++-lambda-capturing mechanism to offload a kernel to GPUs using
CUDA (for NVIDIA GPUs) or HIP (for AMD GPUs).  In our adaptive mesh
hierarchy, each grid is assigned to an MPI task / GPU, and then each
of the zones in that grid are mapped to a single GPU compute core.
This exposes a lot of parallelism and allows us to run an
order-of-magnitude faster using GPUs vs.\ CPUs on a typical
supercomputer node.

\begin{figure}[t]
\centering
\includegraphics[width=0.45\linewidth]{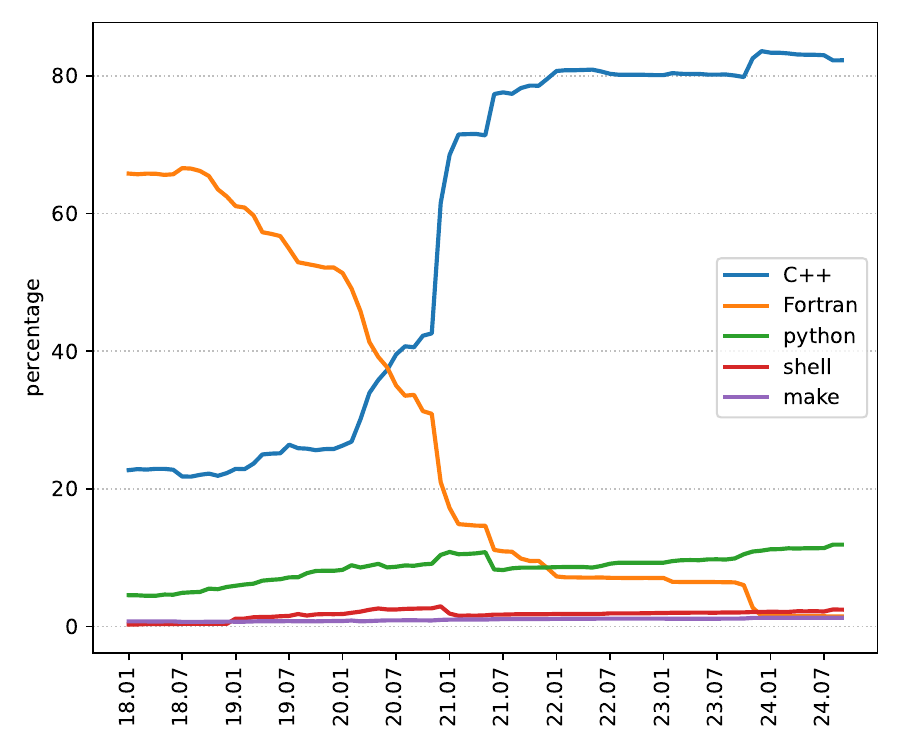}
\includegraphics[width=0.45\linewidth]{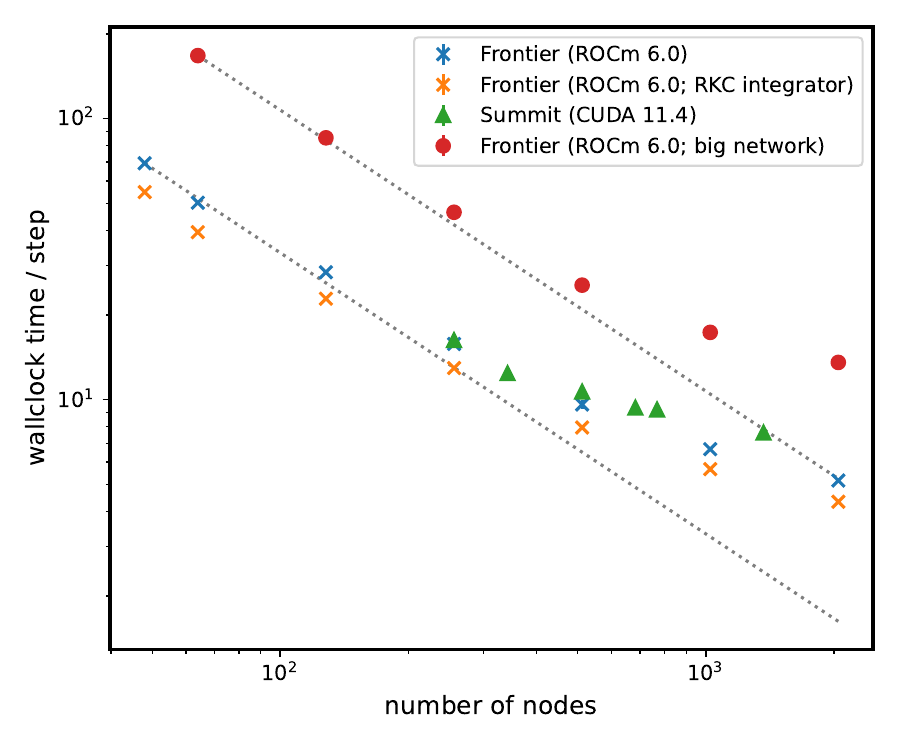}
\caption{\label{fig:castro} (left) the evolution of the programming languages used in \castro,
  showing the sharp decline in Fortran to the fully C++ codebase today (right) strong
  scaling for a 3D XRB flame (from \cite{castro-3d-xrb}) on the OLCF Summit (6 GPUs / node) and Frontier (8 GPUs / node) machines.
  We see good scaling to $\mathcal{O}(10^3)$ nodes.}
\end{figure}

\castro\footnote{\url{https://github.com/amrex-astro/Castro/}} was originally written in a mixed of C++ and Fortran (with the
all of the compute kernels written in Fortran).  To enable this
performance-portable approach to GPUs, we took on the task of porting
the code to C++.  Figure~\ref{fig:castro} shows the history of this transition, where
were were able to move pieces of the code at a time to C++ while
maintaining a functioning code.  This porting paid off, with \castro\
able to strong-scale to $\mathcal{O}(10^3)$ nodes on supercomputers, as shown
in the second panel of Figure~\ref{fig:castro}.  For reactions, our
GPU approach means that the entire ODE integration is managed on GPUs.
This can cause issues with thread-divergence if one zone takes
considerably more time than others---we will discuss some ways of
mitigating this below.

Significant effort has been made by our group to accurately model reactive flows in
stars, going beyond the standard operator-splitting techniques widely
used in the field.  We can write the equations of reactive flow as a 
system in conservative form:
\begin{equation}
    \frac{\partial \Uc}{\partial t} = \Advs{\Uc} + \Rb(\Uc)
\end{equation}
where $\Uc$ represents the conserved state of the fluid (density,
momentum, energy), $\Advs{\Uc}$ is the hydrodynamic advection term, and
$\Rb(\Uc)$ are the reactive sources.  In an operator-split approach to
coupling reactions and hydrodynamics, we treat the advection and
reactions independent of one-another.  In \castro, the $\Advs{\Uc}$ term is
computed using a unsplit piecewise parabolic method
\cite{ppmunsplit,ppm}.  The splitting means that when we integrate the
reaction term, there is no energy / temperature change due to
hydrodynamic expansion / compression.  In flows where the burning is
strongly influencing the flow, like detonations, this means that we
can lose order-of-accuracy or get the nucleosynthesis wrong unless we
cut the timestep (see, e.g., \cite{strang_rnaas,castro_simple_sdc}).

Recently we introduced the ``simplified-SDC'' time-integration method
in \castro~\cite{castro_simple_sdc}, based on the traditional spectral
deferred corrections approach
\cite{dutt:2000,bourlioux:2003,castro-sdc} that we explored earlier.
The simplified SDC method updates the conserved state by evolving
the reaction ODEs with an explicit advection term:
\begin{equation}
\frac{d\Uc}{dt} = \Adv{\Uc}^{n+1/2} + \Rb(\Uc)
\end{equation}
where the advective term, $\Adv{\Uc}^{n+1/2}$, is constructed at the midpoint in time using
characteristic tracing and treated as piecewise-constant-in-time during
the reaction evolution.  
The result is that reactions explicitly sees what advection does over
a timestep.  This allows us to integrate explosive reacting flows
without having to cut the timestep below the hydrodynamics timestep.
All of these time-integration methods are built into
\castro.

The \amrex-Astrophysics codes also share a common set of microphysics
(which we refer to as \amrex-Astro \microphysics\footnote{\url{https://github.com/amrex-astro/Microphysics}}).  This provides
equations of state, reaction networks, transport coefficients, and integrators that can handle
both operator-split and SDC time-integration methods.  For reaction networks, we are tightly integrated with
\pynucastro \cite{pynucastro2}, which allows us to generate a custom
C++ reaction network for any science application with just a few lines of
python.  All of the networks and integrators are written in a manner
that they can be easily offloaded to GPUs.

\section{Confronting Common Practice with Simulations}

In the next sections, we describe some of the choices commonly made in simulations
of astrophysical reacting flows and look at whether alternatives to the standard approaches
can be viable in some situations.

\subsection{Choice of integrator for stiff reaction networks}

Reaction networks are evolved by solving a system of ordinary
differential equations (ODEs), with the integrator taking many small
timesteps to achieve the desired output time.  Historically for
astrophysical reaction networks, simple
backward-Euler or Crank-Nicolson integrators have been used
\cite{arnetttruran:1969,woosley:1973,prometheus,hixmeyer:2006}.
Ideally, the timestep will be controlled by specifying integrator
tolerances and monitoring truncation error, but sometimes heuristics
are used \cite{hixmeyer:2006}.  An extensive study of integrators (and
linear algebra packages) was done by \cite{timmes:1999} and explored a
4th-order Kaps-Rentrop method and a variable order Bader-Deuflhard
method, both of which provide error estimates, and found that the
higher order methods are more efficient and accurate.  These are the
basis for the networks in \flash~\cite{flash}.  The benefits
of high-order integration methods were also explored in 
\cite{longland:2014}, which also concluded that they perform
better than low order methods on a variety of problems.   In our \amrex-Astro \microphysics\ package,
we use the variable order, backward difference implicit VODE
integrator \cite{vode} as the default.  VODE is widely used in the
chemical combustion community (e.g.\ PeleC~\cite{PeleC}), and a C port
is also included in the SUNDIALS library~\cite{sundials}. In
\microphysics, we ported the original VODE Fortran 77 source to a
templated C++ header library, and added some special step rejection
logic that knows about mass fractions (see \cite{castro_simple_sdc}).
This port allows us to offload the entire ODE integration onto GPUs.

All of these implicit integrators require a Jacobian, which is either
evaluated analytically or via finite-differencing.  
Carrying the Jacobian takes
memory (sparse storage helps \cite{timmes:1999}) and expensive linear
algebra operations are needed to advance the state.  
Sometimes Jacobian caching is used, where the same Jacobian
may be reused for multiple steps (if the change in solution
is small) to reduce the computational cost.  However, we
disable Jacobian caching when running on GPUs due to the 
additional memory needed.  The
analytic Jacobian may be approximated, for instance, by ignoring the
composition dependence in the plasma state for screening or in
compound approximation rates (like those used in {\tt aprox19} \cite{Kepler}).
Sometimes the approximate analytic Jacobian works well, but when it
doesn't, the integrator can fail.  In these instances, the numerical
Jacobian often works.  In \microphysics,
we allow for a zone to retry the burn if it fails, swapping the
Jacobian type for the retry.  This proves to be surprisingly robust.

At high temperatures, forward and reverse reaction rates can balance
and the nuclei enter into nuclear statistical equilibrium (NSE).  This
can be challenging for integrators to model, since the forward and
reverse rates can both be large and need to cancel out to high
precision in order to maintain equilibrium.  Integrators that focus on
the quasi-steady state behavior \cite{Mott:2000,Guidry:2013} can help
in some situations, but do not seem to have been widely adopted in
astrophysics.  Methods which detect when the reactive flow enters into
NSE have been used for thermonuclear supernovae
\cite{kushnirkatz:2020} as have heuristic-based approaches to assuming
NSE \cite{ma:2013,bravo:2018}.  We have both approaches implemented in
\microphysics.

\subsubsection{Example: massive stars and nuclear statistical equilibrium.}

For the evolution of the iron core in massive stars
nearing core-collapse, we want to
understand the NSE state from 100s of nuclei, including
electron/positron captures/decays.  This can be accomplished by
tabulating the NSE state and electron fraction evolution in terms of
density, temperature, and electron fraction, and using this NSE state
to compute the energy release and evolution.  NSE tables have been
used with operator splitting for thermonuclear supernova for some time \cite{townsley:2007,seitenzhal:2009,ma:2013}.  In the simplified-SDC
framework, it is possible to capture the time-evolution of the NSE state in a second-order accurate
fashion, using a predictor-corrector approach, as we demonstrated in
\cite{sdc-nse}.

\begin{figure}[t]
\centering
\includegraphics[width=0.8\linewidth]{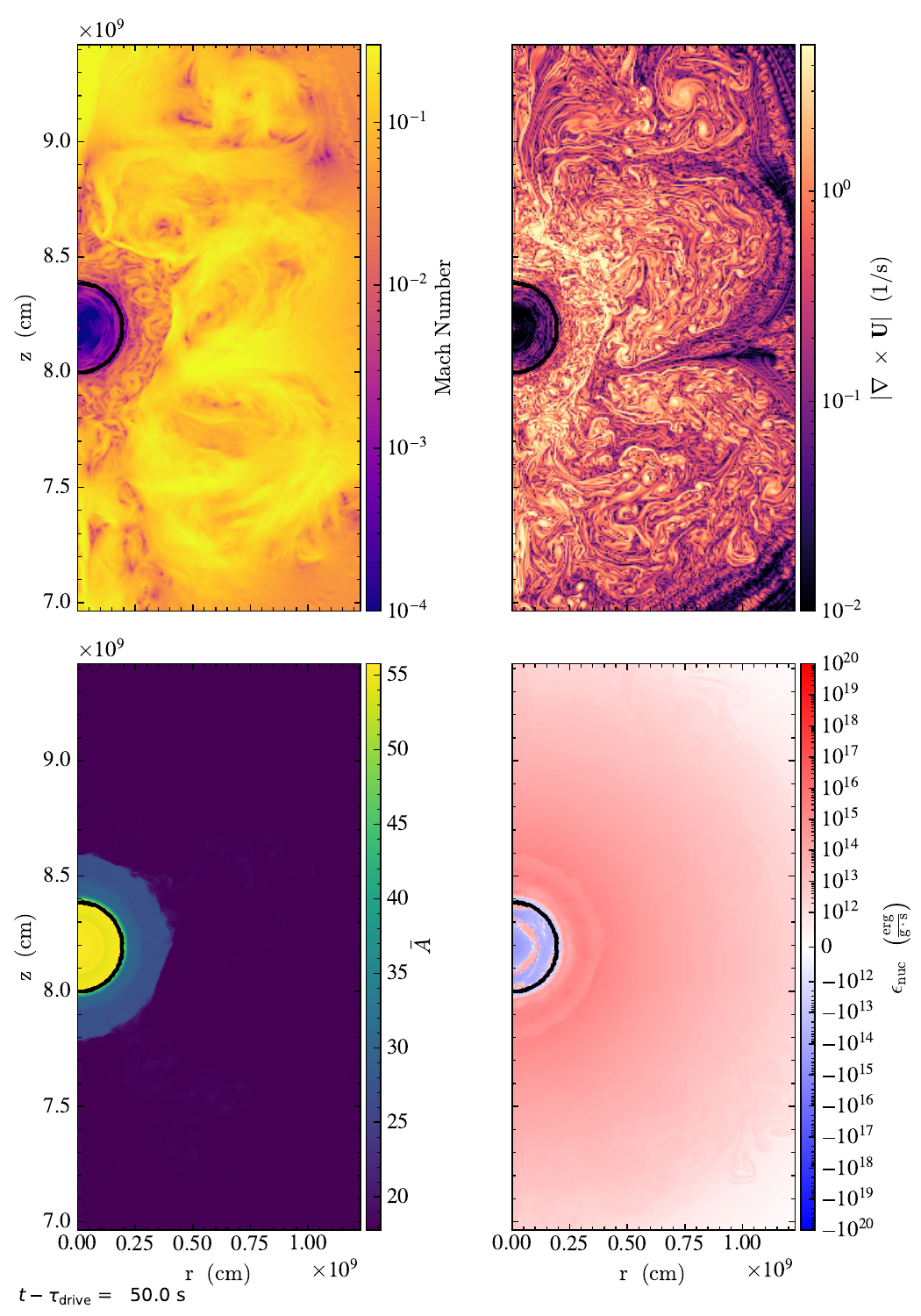}
\caption{\label{fig:massivestar} Mach number, vorticity,
mean molecular weight, and energy generation rate for a $15~M_\odot$
star in the inner
15\% of the simulation domain.  The black contour in each panel
shows the boundary between when we assume NSE and when we integrate
the network.}
\end{figure}

Figure~\ref{fig:massivestar} shows a snapshot of the convective flow
field near the core of a massive star ($15~M_\odot$).  As a fluid
element enters the core, we switch from integrating a 19-isotope
network to using the NSE table to give us the evolution of the electron fraction of the state (due to weak reactions).  The NSE table also maps the $\mathcal{O}(100)$ nuclei to the 19 we carry on the grid,
allowing us continue to track the fluid element if it leaves the core.
With the simplified-SDC time-integration approach, we are able to evolve on the hydrodynamic timestep, without needing to cut the timestep to the much smaller nuclear timestep (as is done in other work).  Furthermore, in contrast to most other works which cutout \cite{muller:2017,yadav:2020,varmamueller:2023} or approximate the core \cite{yoshida:2019,yoshida-apj:2021,fields:2020}, the NSE treatment allows us to include the entire core self-consistently.

The burning in the Si layer nearing NSE is difficult and the integrator can require a lot of internal steps.  We found however that if we put a low cap on the number of ODE integration timesteps ($< 5000$), and allow the burn to fail, and then retry with a different Jacobian approximation, the massive star runs $1.7\times$ faster in 3D on the AMD GPUs on OLCF Frontier.

\subsubsection{Example: X-ray bursts and explicit integrators.}

In chemical combustion, \cite{niemeyersung} showed that stabilized
explicit integrators, like Runge-Kutta-Chebyshev (RKC) \cite{rkc}, can be more
efficient than implicit integrators for moderately-stiff reaction networks, especially on GPUs.  These have
the advantage that a Jacobian is not needed---only an estimate of the spectral
radius is required.  We ported the original Fortran implementation
of RKC to C++ in \microphysics, and are able to use it for
any of our networks.  In \cite{johnson:2023},
we looked at the performance of RKC for flames in XRBs and found that in some situations, these flames can be modeled more efficiently with the explicit RKC integrator, compared to VODE.

\begin{figure}[t]
\includegraphics[width=\linewidth]{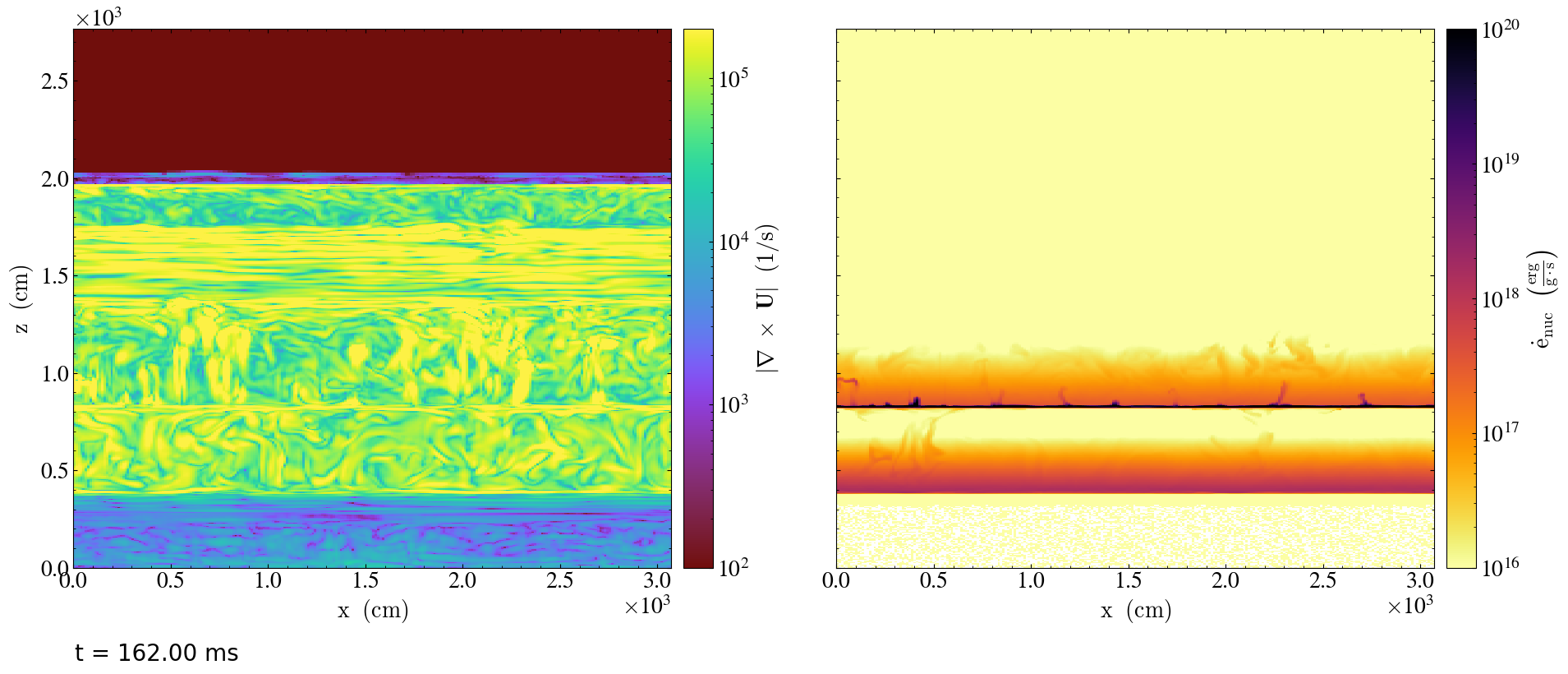}
\caption{\label{fig:xrb_layered} Slice through a 3D simulation of layered
convection in an XRB.  On the left is vorticity, showing the convective field in the main fuel layer.  Only the lower part of the domain is shown---the region above the fuel layer is low density material intended to give the atmosphere room to expand as it heats up.  On the right is the energy generation rate, showing two distinct layers: He burning beneath H burning.}
\end{figure}

The benefits of the explicit integrator are problem-dependent.
Recently we have been exploring ``layered'' models of XRBs (a pure He layer beneath a H layer) \cite{xrb_layered}.  For our latest
3D simulation of layered XRBs (see figure~\ref{fig:xrb_layered}), we found that RKC is almost twice as fast as the implicit VODE integrator with a 21-isotope H-burning network for the initial evolution (as long as the temperature is less than $10^9~\mathrm{K}$).  This suggests that a future optimization
will be to use different integrators for different thermodynamic
conditions in a single simulation.

\subsection{Breakdown of Operator Splitting}

Most astrophysical simulation codes use operator splitting (in particular
Strang splitting \cite{strang:1968}) to couple hydrodynamics and reactions.
This goes back at least to the prometheus code \cite{prometheus}.
Futhermore, the importance of evolving an energy, temperature, or entropy equation
along with the rates was noted by several groups
\cite{muller:1986,prometheus,cabezon:2004}.
In some codes (e.g.\ \cite{flash}), the splitting is further approximated by keeping
the temperature constant during the integration, which we showed can drop the overall
reactive-hydrodynamics evolution to first-order accurate \cite{strang_rnaas}.

Difficulties in coupling hydrodynamics and reactions when doing
operator splitting often requires cutting the timestep.  This is
commonly needed for modeling detonations in the double detonation
progenitor model for Type Ia supernovae (SN Ia).  In our studies of
the double detonation SN Ia problem \cite{castro-sndd}, we found that the
simplified-SDC integration is more accurate than Strang splitting and
also computationally less expensive.  When the reaction integration
includes the advective terms, it seems to remove some of the stiffness
from the problem---we discussed this earlier in \cite{astronum:2019},
where we showed that the simplified-SDC integration evaluates the
righthand side of the reaction ODE system fewer times than the Strang
split case when modeling a detonation.

\subsection{Integration Performance on GPUs}

There are several different approaches to offloading reactions to GPUs.
The Sundials library \cite{sundials} runs the integrator on the CPU
and launches GPU kernels to operate on the state held in GPU memory.
This is done to allow for a batched integration strategy in the
adaptive timestepping algorithm, where a large number of zones are
advanced with the same timestep.  This can reduce thread divergence on
GPUs and could be of benefit for some of our applications.

Our approach matches more with that of \cite{stonedavis:2013}.  We do
the entire ODE integration on the GPU.  Our kernel is the entire VODE
or RKC integrator along with our network-specific righthand side and
Jacobian routines and equation of state.  As discussed with massive stars, one way to help
with thread divergence is to put a cap on the number of ODE
integration steps used by the reaction integrator.  If this fails, then we can
force a retry at the application level, discarding the entire AMR level advance
and redoing it with a smaller timestep.

\section{Summary}

We discussed some recent developments in modeling reactive flows in stars.  For many problems, exploring alternate time-integration strategies that traditionally done in astrophysics can lead to better performance and more accuracy.  All of the developments we discuss here are
freely available in the \amrex-Astro code github repositories.

\ack

\castro\ and the \amrex-Astrophysics suite are freely available at
\url{http://github.com/AMReX-Astro/}.  The work at Stony Brook was
supported by DOE/Office of Nuclear Physics grant DE-FG02-87ER40317
and a seed-funding award from the Stony Brook University Office of
the Vice President for Research.

This research used resources of the National Energy Research
Scientific Computing Center (NERSC), a Department of Energy Office of
Science User Facility using NERSC award NP-ERCAP0027167.  This
research used resources of the Oak Ridge Leadership Computing Facility
at the Oak Ridge National Laboratory, which is supported by the Office
of Science of the U.S. Department of Energy under Contract
No. DE-AC05-00OR22725.

\bibliographystyle{iopart-num}
\bibliography{ws}

\providecommand{\newblock}{}
\begin{thebibliography}{10}
\expandafter\ifx\csname url\endcsname\relax
  \def\url#1{{\tt #1}}\fi
\expandafter\ifx\csname urlprefix\endcsname\relax\def\urlprefix{URL }\fi
\providecommand{\eprint}[2][]{\url{#2}}

\bibitem{sportisse:2000}
Sportisse B 2000 {\em Journal of Computational Physics\/} {\bf 161} 140--168
  ISSN 0021-9991
  \urlprefix\url{https://www.sciencedirect.com/science/article/pii/S0021999100964957}

\bibitem{zingale:2001}
{Zingale} M, {Timmes} F~X, {Fryxell} B, {Lamb} D~Q, {Olson} K, {Calder} A~C,
  {Dursi} L~J, {Ricker} P, {Rosner} R, {MacNeice} P and {Tufo} H~M 2001 {\em
  Astrophysical Journal Supplement\/} {\bf 133} 195--220

\bibitem{Fields_2020}
Fields C~E and Couch S~M 2020 {\em The Astrophysical Journal\/} {\bf 901} 33
  \urlprefix\url{https://dx.doi.org/10.3847/1538-4357/abada7}

\bibitem{rivas:2022}
{Rivas} F, {Harris} J~A, {Hix} W~R and {Messer} O~E~B 2022 {\em arXiv
  e-prints\/} arXiv:2205.12370 (\textit{Preprint} \eprint{2205.12370})

\bibitem{astronum:2017}
Zingale M, Almgren A~S, Barrios~Sazo M~G, Beckner V~E, Bell J~B, Friesen B,
  Jacobs A~M, Katz M~P, Malone C~M, Nonaka A~J, Willcox D~E and Zhang W 2018
  {\em J. Phys.: Conf. Ser.\/} {\bf 1031} 012024 ISSN 1742-6588, 1742-6596
  (\textit{Preprint} \eprint{1711.06203})
  \urlprefix\url{https://doi.org/10.1088/1742-6596/1031/1/012024}

\bibitem{amrex_joss}
Zhang W, Almgren A, Beckner V, Bell J, Blaschke J, Chan C, Day M, Friesen B,
  Gott K, Graves D, Katz M, Myers A, Nguyen T, Nonaka A, Rosso M, Williams S
  and Zingale M 2019 {\em JOSS\/} {\bf 4} 1370 ISSN 2475-9066
  \urlprefix\url{https://doi.org/10.21105/joss.01370}

\bibitem{castro}
Almgren A~S, Beckner V~E, Bell J~B, Day M~S, Howell L~H, Joggerst C~C, Lijewski
  M~J, Nonaka A, Singer M and Zingale M 2010 {\em ApJ\/} {\bf 715} 1221--1238
  ISSN 0004-637X, 1538-4357 (\textit{Preprint} \eprint{1005.0114})
  \urlprefix\url{https://doi.org/10.1088/0004-637x/715/2/1221}

\bibitem{castro_joss}
Almgren A, Sazo M~B, Bell J, Harpole A, Katz M, Sexton J, Willcox D, Zhang W
  and Zingale M 2020 {\em Journal of Open Source Software\/} {\bf 5} 2513
  \urlprefix\url{https://doi.org/10.21105/joss.02513}

\bibitem{maestroex}
{Fan} D, {Nonaka} A, {Almgren} A~S, {Harpole} A and {Zingale} M 2019 {\em
  Astrophysical Journal\/} {\bf 887} 212 (\textit{Preprint}
  \eprint{1908.03634})

\bibitem{nyx}
{Almgren} A~S, {Bell} J~B, {Lijewski} M~J, {Luki{\'c}} Z and {Van Andel} E 2013
  {\em Astrophysical Journal\/} {\bf 765} 39 (\textit{Preprint}
  \eprint{1301.4498})

\bibitem{castro_gpu}
Katz M~P, Almgren A, Sazo M~B, Eiden K, Gott K, Harpole A, Sexton J~M, Willcox
  D~E, Zhang W and Zingale M 2020 {\em Proceedings of the International
  Conference for High Performance Computing, Networking, Storage and
  Analysis\/} SC '20 (IEEE Press) ISBN 9781728199986
  \urlprefix\url{https://dl.acm.org/doi/abs/10.5555/3433701.3433822}

\bibitem{castro-3d-xrb}
Zingale M, Eiden K and Katz M 2023 {\em The Astrophysical Journal\/} {\bf 952}
  160 \urlprefix\url{https://dx.doi.org/10.3847/1538-4357/ace04e}

\bibitem{ppmunsplit}
Colella P 1990 {\em J. Comput. Phys.\/} {\bf 87} 171--200 ISSN 0021-9991
  \urlprefix\url{https://doi.org/10.1016/0021-9991(90)90233-q}

\bibitem{ppm}
Colella P and Woodward P~R 1984 {\em J. Comput. Phys.\/} {\bf 54} 174--201 ISSN
  0021-9991 \urlprefix\url{https://doi.org/10.1016/0021-9991(84)90143-8}

\bibitem{strang_rnaas}
Zingale M, Katz M~P, Willcox D~E and Harpole A 2021 {\em Research Notes of the
  AAS\/} {\bf 5} 71 \urlprefix\url{https://dx.doi.org/10.3847/2515-5172/abf3cb}

\bibitem{castro_simple_sdc}
Zingale M, Katz M~P, Nonaka A and Rasmussen M 2022 {\em The Astrophysical
  Journal\/} {\bf 936} 6
  \urlprefix\url{https://dx.doi.org/10.3847/1538-4357/ac8478}

\bibitem{dutt:2000}
{Dutt} A, {Greengard} L and {Rokhlin} V 2000 {\em BIT Numerical Mathematics\/}
  {\bf 40}(2) 241--266

\bibitem{bourlioux:2003}
Bourlioux A, Layton A~T and Minion M~L 2003 {\em Journal of Computational
  Physics\/} {\bf 189} 651--675

\bibitem{castro-sdc}
Zingale M, Katz M~P, Bell J~B, Minion M~L, Nonaka A~J and Zhang W 2019 {\em
  ApJ\/} {\bf 886} arXiv:1908.03661 ISSN 1538-4357 accepted to ApJ
  (\textit{Preprint} \eprint{1908.03661})
  \urlprefix\url{https://doi.org/10.3847/1538-4357/ab4e1d}

\bibitem{pynucastro2}
Smith A~I, Johnson E~T, Chen Z, Eiden K, Willcox D~E, Boyd B, Cao L, DeGrendele
  C~J and Zingale M 2023 {\em The Astrophysical Journal\/} {\bf 947} 65
  \urlprefix\url{https://dx.doi.org/10.3847/1538-4357/acbaff}

\bibitem{arnetttruran:1969}
{Arnett} W~D and {Truran} J~W 1969 {\em Astrophysical Journal\/} {\bf 157} 339

\bibitem{woosley:1973}
{Woosley} S~E, {Arnett} W~D and {Clayton} D~D 1973 {\em Astrophysical Journal
  Supplement Series\/} {\bf 26} 231

\bibitem{prometheus}
{Fryxell} B~A, {M{\"u}ller} E and {Arnett} D 1989 {\em MPA Preprint 449\/}

\bibitem{hixmeyer:2006}
Hix W~R and Meyer B~S 2006 {\em Nuclear Physics A\/} {\bf 777} 188--207 ISSN
  0375-9474 special Isseu on Nuclear Astrophysics
  \urlprefix\url{https://www.sciencedirect.com/science/article/pii/S0375947404011005}

\bibitem{timmes:1999}
Timmes F~X 1999 {\em The Astrophysical Journal Supplement Series\/} {\bf 124}
  241 \urlprefix\url{https://dx.doi.org/10.1086/313257}

\bibitem{flash}
{Fryxell} B, {Olson} K, {Ricker} P, {Timmes} F~X, {Zingale} M, {Lamb} D~Q,
  {MacNeice} P, {Rosner} R, {Truran} J~W and {Tufo} H 2000 {\em Astrophysical
  Journal Supplement Series\/} {\bf 131} 273{\textendash}334

\bibitem{longland:2014}
{Longland} R, {Martin} D and {Jos{\'e}} J 2014 {\em Astronomy and
  Astrophysics\/} {\bf 563} A67 (\textit{Preprint} \eprint{1401.5762})

\bibitem{vode}
Brown P~N, Byrne G~D and Hindmarsh A~C 1989 {\em SIAM J. Sci. and Stat.
  Comput.\/} {\bf 10} 1038--1051 ISSN 0196-5204, 2168-3417
  \urlprefix\url{https://doi.org/10.1137/0910062}

\bibitem{PeleC}
{Henry de Frahan} M~T, Rood J~S, Day M~S, Sitaraman H, Yellapantula S, Perry
  B~A, Grout R~W, Almgren A, Zhang W, Bell J~B and Chen J~H 2022 {\em The
  International Journal of High Performance Computing Applications\/} {\bf 37}
  115--131 \urlprefix\url{https://doi.org/10.1177/10943420221121151}

\bibitem{sundials}
{Balos} C~J, {Day} M, {Esclapez} L, {Felden} A~M, {Gardner} D~J, {Hassanaly} M,
  {Reynolds} D~R, {Rood} J, {Sexton} J~M, {Wimer} N~T and {Woodward} C~S 2024
  {\em arXiv e-prints\/} arXiv:2405.01713 (\textit{Preprint}
  \eprint{2405.01713})

\bibitem{Kepler}
Weaver T~A, Zimmerman G~B and Woosley S~E 1978 {\em ApJ\/} {\bf 225} 1021 ISSN
  0004-637X, 1538-4357 \urlprefix\url{https://doi.org/10.1086/156569}

\bibitem{Mott:2000}
Mott D~R, Oran E~S and {van Leer} B 2000 {\em Journal of Computational
  Physics\/} {\bf 164} 407--428 ISSN 0021-9991
  \urlprefix\url{https://www.sciencedirect.com/science/article/pii/S0021999100966051}

\bibitem{Guidry:2013}
Guidry M~W and Harris J~A 2013 {\em Computational Science \& Discovery\/} {\bf
  6} 015002 \urlprefix\url{https://dx.doi.org/10.1088/1749-4699/6/1/015002}

\bibitem{kushnirkatz:2020}
{Kushnir} D and {Katz} B 2020 {\em Monthly Notices of the Royal Astronomical
  Society\/} {\bf 493} 5413--5433 (\textit{Preprint} \eprint{1912.06151})

\bibitem{ma:2013}
{Ma} H, {Woosley} S~E, {Malone} C~M, {Almgren} A and {Bell} J 2013 {\em
  Astrophysical Journal\/} {\bf 771} 58 (\textit{Preprint} \eprint{1305.2433})

\bibitem{bravo:2018}
Bravo E, Badenes C and Martínez-Rodríguez H 2018 {\em Monthly Notices of the
  Royal Astronomical Society\/} {\bf 482} 4346--4363 ISSN 0035-8711
  (\textit{Preprint}
  \eprint{https://academic.oup.com/mnras/article-pdf/482/4/4346/26779042/sty2951.pdf})
  \urlprefix\url{https://doi.org/10.1093/mnras/sty2951}

\bibitem{townsley:2007}
{Townsley} D~M, {Calder} A~C, {Asida} S~M, {Seitenzahl} I~R, {Peng} F,
  {Vladimirova} N, {Lamb} D~Q and {Truran} J~W 2007 {\em \apj\/} {\bf 668}
  1118--1131 (\textit{Preprint} \eprint{0706.1094})

\bibitem{seitenzhal:2009}
{Seitenzahl} I~R, {Townsley} D~M, {Peng} F and {Truran} J~W 2009 {\em Atomic
  Data and Nuclear Data Tables\/} {\bf 95} 96--114

\bibitem{sdc-nse}
{Zingale} M, {Chen} Z, {Johnson} E~T, {Katz} M~P and {Smith Clark} A 2024 {\em
  arXiv e-prints\/} arXiv:2403.14786 (\textit{Preprint} \eprint{2403.14786})

\bibitem{muller:2017}
{M{\"u}ller} B, {Melson} T, {Heger} A and {Janka} H~T 2017 {\em \mnras\/} {\bf
  472} 491--513 (\textit{Preprint} \eprint{1705.00620})

\bibitem{yadav:2020}
{Yadav} N, {M{\"u}ller} B, {Janka} H~T, {Melson} T and {Heger} A 2020 {\em
  \apj\/} {\bf 890} 94 (\textit{Preprint} \eprint{1905.04378})

\bibitem{varmamueller:2023}
{Varma} V and {M{\"u}ller} B 2023 {\em \mnras\/} {\bf 526} 5249--5262
  (\textit{Preprint} \eprint{2307.04833})

\bibitem{yoshida:2019}
Yoshida T, Takiwaki T, Kotake K, Takahashi K, Nakamura K and Umeda H 2019 {\em
  The Astrophysical Journal\/} {\bf 881} 16
  \urlprefix\url{https://dx.doi.org/10.3847/1538-4357/ab2b9d}

\bibitem{yoshida-apj:2021}
Yoshida T, Takiwaki T, Kotake K, Takahashi K, Nakamura K and Umeda H 2021 {\em
  The Astrophysical Journal\/} {\bf 908} 44
  \urlprefix\url{https://dx.doi.org/10.3847/1538-4357/abd3a3}

\bibitem{fields:2020}
{Fields} C~E and {Couch} S~M 2020 {\em \apj\/} {\bf 901} 33 (\textit{Preprint}
  \eprint{2008.04266})

\bibitem{niemeyersung}
Niemeyer K~E and Sung C~J 2014 {\em Journal of Computational Physics\/} {\bf
  256} 854--871 ISSN 0021-9991
  \urlprefix\url{https://www.sciencedirect.com/science/article/pii/S0021999113006396}

\bibitem{rkc}
Sommeijer B, Shampine L and Verwer J 1998 {\em Journal of Computational and
  Applied Mathematics\/} {\bf 88} 315--326 ISSN 0377-0427
  \urlprefix\url{https://www.sciencedirect.com/science/article/pii/S0377042797002197}

\bibitem{johnson:2023}
Johnson P, Zingale M, Johnson E~T, Smith A and Niemeyer K~E 2023 {\em Research
  Notes of the AAS\/} {\bf 7} 282
  \urlprefix\url{https://dx.doi.org/10.3847/2515-5172/ad175d}

\bibitem{xrb_layered}
Guichandut S, Zingale M and Cumming A 2024 {\em The Astrophysical Journal\/}
  {\bf 975} 250 \urlprefix\url{https://dx.doi.org/10.3847/1538-4357/ad81f7}

\bibitem{strang:1968}
Strang G 1968 {\em SIAM J. Numer. Anal.\/} {\bf 5} 506--517 ISSN 0036-1429,
  1095-7170 \urlprefix\url{https://doi.org/10.1137/0705041}

\bibitem{muller:1986}
{M{\"u}ller} E 1986 {\em Astron. Astrophys.\/} {\bf 162} 103--108

\bibitem{cabezon:2004}
Cabezón R~M, García-Senz D and Bravo E 2004 {\em The Astrophysical Journal
  Supplement Series\/} {\bf 151} 345
  \urlprefix\url{https://dx.doi.org/10.1086/382352}

\bibitem{castro-sndd}
Zingale M, Chen Z, Rasmussen M, Polin A, Katz M, Clark A~S and Johnson E~T 2024
  {\em The Astrophysical Journal\/} {\bf 966} 150
  \urlprefix\url{https://dx.doi.org/10.3847/1538-4357/ad3441}

\bibitem{astronum:2019}
Zingale M, Eiden K, Cavecchi Y, Harpole A, Bell J~B, Chang M, Hawke I, Katz
  M~P, Malone C~M, Nonaka A~J, Willcox D~E and Zhang W 2019 {\em Journal of
  Physics: Conference Series\/} {\bf 1225} 012005
  \urlprefix\url{https://dx.doi.org/10.1088/1742-6596/1225/1/012005}

\bibitem{stonedavis:2013}
Stone C~P and Davis R~L 2013 {\em Journal of Propulsion and Power\/} {\bf 29}
  764--773 (\textit{Preprint} \eprint{https://doi.org/10.2514/1.B34874})
  \urlprefix\url{https://doi.org/10.2514/1.B34874}

\end{thebibliography}

\end{document}